\documentclass[]{spie}  

\usepackage{amsmath,amsfonts,amssymb}
\usepackage{graphicx}
\usepackage[colorlinks=true, allcolors=blue]{hyperref}

\title{Addressing the low-wind effect with the Lyot-based low-order wavefront sensor on SCExAO}

\author[a,b]{Garima Singh}
\author[a,c,d]{S\'ebastien Vievard}
\author[a,b]{Julien Lozi}
\author[a,e]{Vincent Deo}
\author[f]{Andre Fogal}
\author[a,g]{Kyohoon Anh}
\author[a,b,h,i]{Olivier Guyon}
\author[a,h]{Sandrine Juillard}
\affil[a]{National Astronomical Observatory of Japan, Subaru Telescope, 650 North Aohoku Place, Hilo, HI 96720, USA}
\affil[b]{Astrobiology Center, 2-21-1, Osawa, Mitaka, Tokyo 181-8588, Japan}
\affil[c]{Space Science and Engineering Initiative, College of Engineering, University of Hawai`i, Hilo, HI 96720, USA}
\affil[d]{Institute for Astronomy, University of Hawaii, Hilo, HI 96720, USA}
\affil[e]{Optical Sharpeners, Manosque, France}
\affil[f]{University of Victoria, 3800 Finnerty Road, Victoria, BC, V8P 5C2, Canada}
\affil[g]{Korea Astronomy and Space Science Institute (KASI), Daejeon 34055, Republic of Korea}
\affil[h]{Steward Observatory, University of Arizona, Tucson, AZ 87521, United States}
\affil[i]{College of Optical Sciences, University of Arizona, Tucson, AZ 85721, USA}

\authorinfo{Send correspondence to E-mail: gsingh@naoj.org}

\begin{document} 
\maketitle

\begin{abstract}
The Low-Wind Effect (LWE) is a well-known issue that affects the coronagraphic capabilities of instruments optimized to take direct images of exoplanets. This effect is prominent in segmented and obstructed pupils with spider arms, causing differential pistons due to phase discontinuities and, in some cases, tip-tilt errors under low-wind conditions. LWE-induced low-order errors contribute to coronagraphic leakage and reduce the exoplanet detection sensitivity of direct imaging instruments.

Solutions to mitigate the impact of LWE have been developed in both passive mode, such as coating the spider arms with low-emissivity material, and active mode using wavefront sensors (WFS). The Subaru Coronagraphic Extreme Adaptive Optics (SCExAO) instrument at the Subaru Telescope also attempted to mitigate LWE by testing several dedicated WFS. Such techniques, when tested on SCExAO, have proven their ability to measure and correct LWE; however, most remain incompatible with SCExAO's coronagraphic imaging modes.

Addressing the need for an efficient solution that can correct LWE in coronagraphic observing modes, we revisited the coronagraphic WFS on SCExAO, known as the Lyot-based Low-order Wavefront Sensor (LLOWFS). LLOWFS utilizes the unused starlight reflected off the Lyot stop and is a well-proven technique that can prevent coronagraphic leaks by sensing low-order errors in a re-imaged focal plane or pupil plane downstream of the focal plane mask.

We present the machine-learned LLOWFS's initial on-sky measurement and control of the differential piston aberrations induced by the LWE downstream of a Lyot coronagraph in SCExAO's infrared arm. Preliminary LLOWFS corrections of LWE in the coronagraphic mode are encouraging and would leverage future high-contrast performance of SCExAO, enabling the detection of mature exoplanets at small angular separations.   
\end{abstract}


\section{INTRODUCTION}
\label{sec:intro} 
A common scientific goal of current state-of-the-art High-Contrast Imaging (HCI) instruments is to enable imaging of mature planets at solar-system distances. The main challenge preventing such detections is the uncalibrated residual speckle field at small angular resolution. One source of residual starlight near the small Inner Working Angle (IWA) of coronagraphs is starlight leakage due to low-order wavefront aberrations caused by thermal distortion, pointing errors, optical misalignment, and vibration. The SPHERE\cite{Beuzit06} instrument at VLT became the first HCI instrument to introduce yet another source of starlight leakage, called the Low-Wind Effect\cite{lwe-1949, lwe-2015, lwe-2018} (LWE), caused by a temperature gradient between the spider arms of the telescope and the surrounding air for wind speeds below 3~m/s. This effect is well-known to introduce differential piston, tip, and tilt phase errors between disconnected regions of the pupil. It remains unseen by the Pyramid or Shack-Hartmann wavefront sensors because these sensors measure local phase gradients rather than phase discontinuities across the telescope spiders. 

The Subaru Coronagraphic Extreme Adaptive Optics (SCExAO\cite{nem_scexao}) instrument at the Subaru Telescope also experiences LWE, with 30–50\% of SCExAO's science nights being affected. Subaru is equipped with its facility Adaptive Optics (AO) system, AO3K, which comprises a 3228-actuator deformable mirror (DM) and a near-infrared (NIR) Pyramid wavefront sensor. AO3K feeds SCExAO, which incorporates a 2000-actuator DM and a visible Pyramid wavefront sensor. Together, these systems achieve on-sky Strehl ratios exceeding 92\% in the H band\cite{ao3k-lozi}. CHARIS\cite{charis}, which is an Integral Field Spectrograph, is the main science instrument for AO3k-SCExAO. Despite two levels of Extreme-AO correction at Subaru, the effects of low-wind modes remain uncorrected. Several techniques\cite{ao3k-lozi} have been implemented on SCExAO to correct for the LWE, and the Fast and Furious\cite{FnF} technique has also been tested on-sky; however, these techniques only works in non-coronagraphic mode. In this work, we present a machine learning approach with the Lyot-based Low-Order Wavefront Sensor (LLOWFS\cite{singh1}) on SCExAO to tackle differential piston caused by LWE in coronagraphic mode. Sect.~\ref{sec:llowfs} introduces LLOWFS and the machine learning method tested on SCExAO, and Sect.~\ref{sec:results} presents preliminary laboratory and on-sky results.    

\section{Lyot-based Low-order Wavefront Sensor}
\label{sec:llowfs} 

LLOWFS is an established coronagraphic low-order wavefront sensor that provides efficient low-order wavefront control for low IWA phase-mask coronagraphs\cite{singh_onsky, singh3}. LLOWFS, located on the NIR arm of SCExAO, re-images the on-axis starlight diffracted by the focal-plane mask at the Lyot plane into a separate channel, using this near-focal-plane signal for low-order wavefront sensing. The optical layout of LLOWFS, its very first simulation, laboratory implementation, and detailed on-sky performance on SCExAO at the Subaru Telescope can be found in Singh et al \cite{singh1, singh_onsky, singh3}. 

LLOWFS, also referred to as Lyot-LOWFS or Lyot-stop LOWFS in the HCI community, has been widely adopted by independent groups internationally, such as MagAO-X\cite{magao-llowfs} at Magellan in Chile, Canada’s SPIDERS\cite{spiders-llowfs} at Subaru, CAL2\cite{cal2-llowfs} for GPI2.0 at Gemini North, the University of Massachusetts Lowell and NASA’s PICTURE-C\cite{picutreC-llowfs} and D\cite{picutreD-llowfs} balloon missions, the University of Arizona’s space coronagraph optical bench\cite{scoob-llowfs}, Paris Observatory's THD2\cite{thd2-llowfs} bench, and subsequently suggested for future  ground-\cite{gmt-llowfs} and space-based\cite{space-llowfs2, space-llowfs1} HCI instruments.

Classical LLOWFS, despite its efficient wavefront control for low-order errors, has a limited linear range within $\pm100$ nm rms and becomes unstable in large aberration regimes. Allan et al.\cite{allan-ml-llowfs} implemented a Convolutional Neural Network (CNN) architecture based on ResNet\cite{resnet} with LLOWFS and demonstrated an increase in its dynamic range for 15 Zernike modes. Their simulated study conducted with a four-quadrant phase-mask coronagraph at a wavelength ($\lambda$) of 635 nm showed a linear sensor response to wavefront error up to 1.5 waves in low-photon regimes. Fogal et al.\cite{fogal-ml-llowfs} recently implemented a CNN approach based on Landman et al.\cite{Landman1} for LLOWFS installed on the SPIDERS instrument, and demonstrated on-sky correction of over 14 low-order modes at Subaru, improving linearity and reducing crosstalk. We implemented a similar CNN architecture as Fogal et al. for LLOWFS on SCExAO to correct for the differential piston aberrations induced by the LWE. In our experiments, we used an opaque FPM with an IWA of 113~milli-arcsecond (mas). Sections~\ref{sec:cnn} and~\ref{sec:results} briefly present LLOWFS's machine learning procedure and the results obtained in the laboratory and on-sky at Subaru, respectively. 

\subsection{Machine-Learning with LLOWFS}
\label{sec:cnn} 
For aberrations smaller than the 1~radian rms wavefront error, the intensity fluctuation in the LLOWFS image is a linear function of the low-order phase errors in the pupil plane upstream of a coronagraph. However, this linear relationship breaks down for large aberrations as a result of higher-order effects that cannot be captured by a linear interaction matrix. Neural networks have been used to reconstruct wavefront in non-linear regimes from measured sensor images by approximating arbitrary functions and solving the inverse problem\cite{Landman1}. We used a CNN as a reconstructor that learned a nonlinear mapping between the low-wind modes at the pupil plane and the corresponding defocused LLOWFS images. We provide below a brief description of the steps followed in generating a CNN reconstructor. The reader may also refer to Fogal et al. for alternative explanations of the method. A detailed description of the procedure followed will be presented in future publications.    

\subsubsection{LLOWFS Training Library}
\label{sec:cnn-lib}
The first step in Machine-Learning (ML) with LLOWFS (ML-LLOWFS) is to generate a supervised training dataset. We experimentally created a library that paired known pupil-quadrant piston modes with their corresponding LLOWFS intensity responses at the wavelength of 1.65$\mu$m on SCExAO. For each sample, we first create a known coefficient vector containing amplitudes for each piston mode:

\[a=(a_{1}, a_{2}, ..., a_{N})^T,\]

where $N$ is the total number of modes, which is four in our case.

\begin{figure} [ht]
\begin{center}
\begin{tabular}{c} 
\includegraphics[height=12cm]{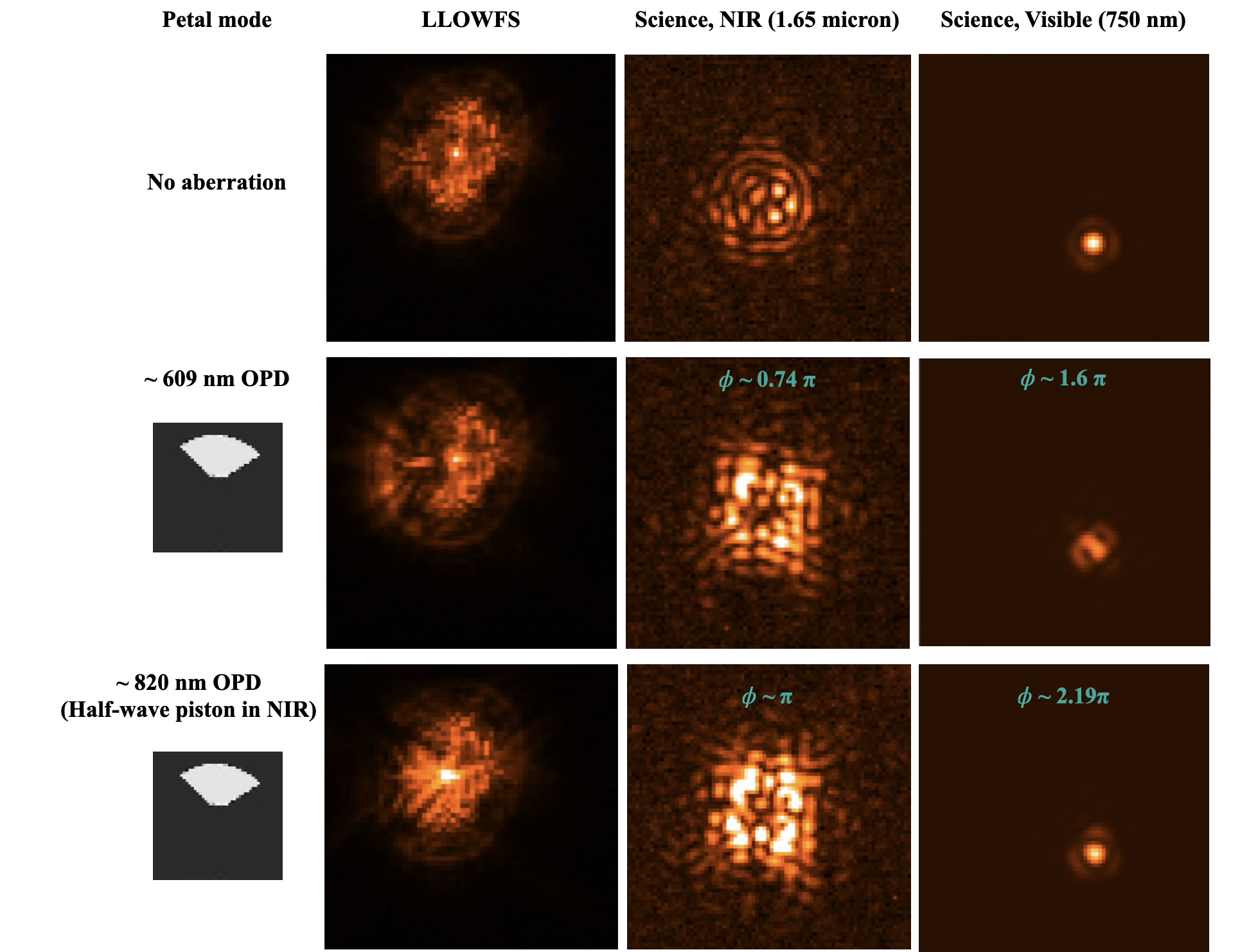}
\end{tabular}
\end{center}
\caption[example] 
{\label{fig:responses} The laboratory response of the SCExAO detectors for a piston mode. The first column shows the piston mode applied on the DM with two different optical path differences. The second, third, and fourth columns show the corresponding LLOWFS, coronagraphic NIR science, and non-coronagraphic visible science images, respectively. The first row shows images with no aberration applied on the DM. An 820~nm piston mode-induced OPD has minimal impact on the visible image because it corresponds to approximately a 2$\pi$ phase shift at 750~nm, whereas it introduces a $\pi$ phase shift at 1.65$\mu$m in the NIR, leading to stronger image degradation. An opaque coronagraphic mask with IWA of 113 mas is used in the NIR arm.}
\end{figure}

We then create a phasemap ($\phi(x,y)$) with a linear combination of four pupil-quadrant piston modes and apply it to the DM:

\[\phi(x,y)=\sum_{j=1}^{N} a_j P_j(x,y),\]

where $P_j(x,y)$ represents the spatial structure of the $j^{th}$ low-wind mode, $a_{j}$ is the corresponding amplitude, and the variables $x$ and $y$ represent the spatial coordinates across the pupil plane. The measured response of LLOWFS to each applied phasemap was recorded as $I_{L}(x,y)$. Each $I_{L}(x,y)$ was an average of 100 images. Figure~\ref{fig:responses} shows a snapshot of the applied piston mode and the corresponding LLOWFS response. We also show the impact of the same piston mode in the visible from the VAMPIRES\cite{vampires} module situated in the visible arm of SCExAO. These images are also shown for two optical path difference (OPD) values depicting wavelength-dependent phase wrapping. An 820~nm OPD corresponds to approximately a 2$\pi$ phase shift at 750~nm, making the aberration nearly invisible in the visible image, whereas it corresponds to approximately a $\pi$ phase shift at 1.65$\mu$m, producing a strong impact in the NIR image. 

We produced 50,000 measurements, each with a supervised training pair: $(I_{L}(x,y), a)$, where $I_{L}$ acts as input to the CNN model under learning (section~\ref{sec:cnn-arch}) and the known coefficients of piston modes, $a$, serve as the verifiable output ($\hat{a}$).    

The amplitude coefficient space $a$ to generate the library containing the linear and non-linear responses of LLOWFS was sampled using a combination of deterministic single-mode scans, random multi-mode aberration combinations, and zero-aberration reference points. For the single-mode scan, the piston mode in each pupil quadrant was consecutively applied to the DM over the range of $\pm5$~rad using 71 amplitude values to characterize the individual sensitivities and the linearity of the LLOWFS response to each mode. The distribution of the amplitude for each mode was non-uniform, with more sampling points within $\pm$1~radian where the sensor is highly linear. 

For the multi-mode scan, random combinations of differential piston modes between the pupil quadrants were included in each coefficient vector. The amplitude value of each mode was drawn from a zero-mean Gaussian distribution:

\[a_j \sim \mathcal{N}(0,\sigma_j^2),\]

using mode-dependent standard deviations

\[\sigma=[1.0,1.5,0.8,1.2].\]

This is to emulate the closed-loop regime under normal operating conditions while also including sufficiently large aberrations to feed nonlinear behavior in the CNN model. We also included data points under the no-aberration ($a = 0$) regime as reference measurements to help stabilize the regression around zero wavefront error. The coefficient vectors were also reshuffled before being applied to the DM to eliminate ordering bias between consecutive measurements. 

\subsubsection{Network Architecture}
\label{sec:cnn-arch}
Once the library is computed, the reconstruction problem becomes 
\[a = f(I_{L}),\]
where $f$ is the unknown nonlinear inverse mapping. The CNN then learns a numerical approximation 

\[\hat{a} = f_{\theta}(I_{L}),\]

where $\theta$ signifies the trainable network parameters. 

Before network training, each $I_{L}$ is normalized, mean-subtracted, and divided by the standard deviation. This enables the CNN model to learn the shape of the LLOWFS intensity pattern rather than changes in intensity due to varying seeing. 

CNN training is based on convolutional layers that transform $I_{L}$ into localized optical signatures such as diffraction features and asymmetric intensity distributions associated with particular wavefront aberrations. To lower computational cost, max-pooling is performed to reduce the spatial dimensions of the feature maps without losing dominant responses, followed by flattening the feature maps into a one-dimensional vector and passing through three fully connected layers.

The network parameters were optimized by minimizing the normalized root mean squared error between the amplitudes predicted by the network, $\hat{a}$, and applied modal coefficients, $a$, as defined by the loss function ($L$) in Landman et al.\cite{Landman1}.

\[
L =
\left\langle\frac{
\sqrt{\sum_{j=1}^{N}(a_j-\hat{a}_j)^2}
}{
\sqrt{\sum_{j=1}^{N}a_j^2}+0.01
}\right\rangle
.\]

\begin{figure} [ht]
\begin{center}
\begin{tabular}{c} 
\includegraphics[height=6cm]{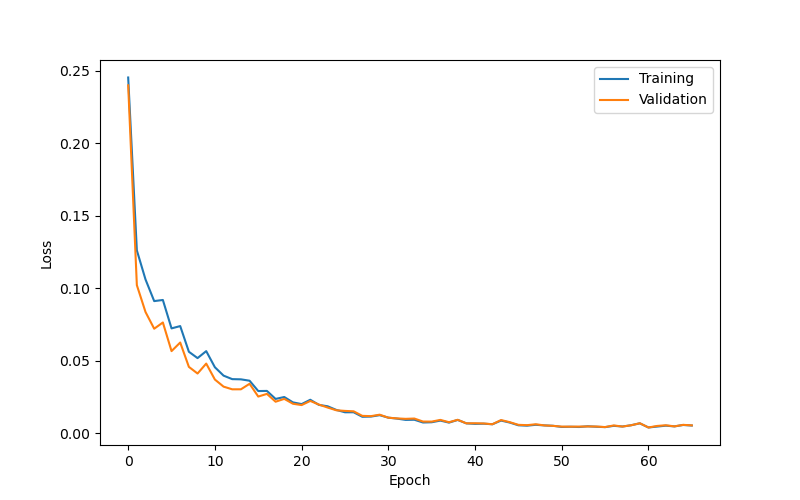}
\end{tabular}
\end{center}
\caption[example] 
{\label{fig:loss} Loss curves for the training and validation datasets. The curves converged beyond epoch 25, suggesting the CNN model is not overfitting.}
\end{figure}

The network was trained for 65 epochs using the training data, with a batch size of 256. The model parameters were optimized using the ADAM\cite{opt} optimizer with a learning rate of $10^{-4}$. The performance of the CNN model was evaluated by randomly splitting the library into 80\% training data and 20\% validation data. At the end of each epoch, the validation loss was computed on the validation dataset using the loss function defined above to assess the CNN model's generalization performance. The training and validation data loss curves are shown in Fig.~\ref{fig:loss}, indicating that the validation loss decreases and converges with the training loss, suggesting that the model is not overfitting. It took about 45 minutes to obtain the library in the laboratory and a few minutes to train the CNN model to predict the amplitudes of the four pupil-quadrant piston modes. Training was performed using CUDA on an NVIDIA RTX A6000 GPU, with 82 CPU threads used for parallel computation.

\section{Results}
\label{sec:results} 
\begin{figure} [ht]
\begin{center}
\begin{tabular}{c} 
\includegraphics[height=12cm]{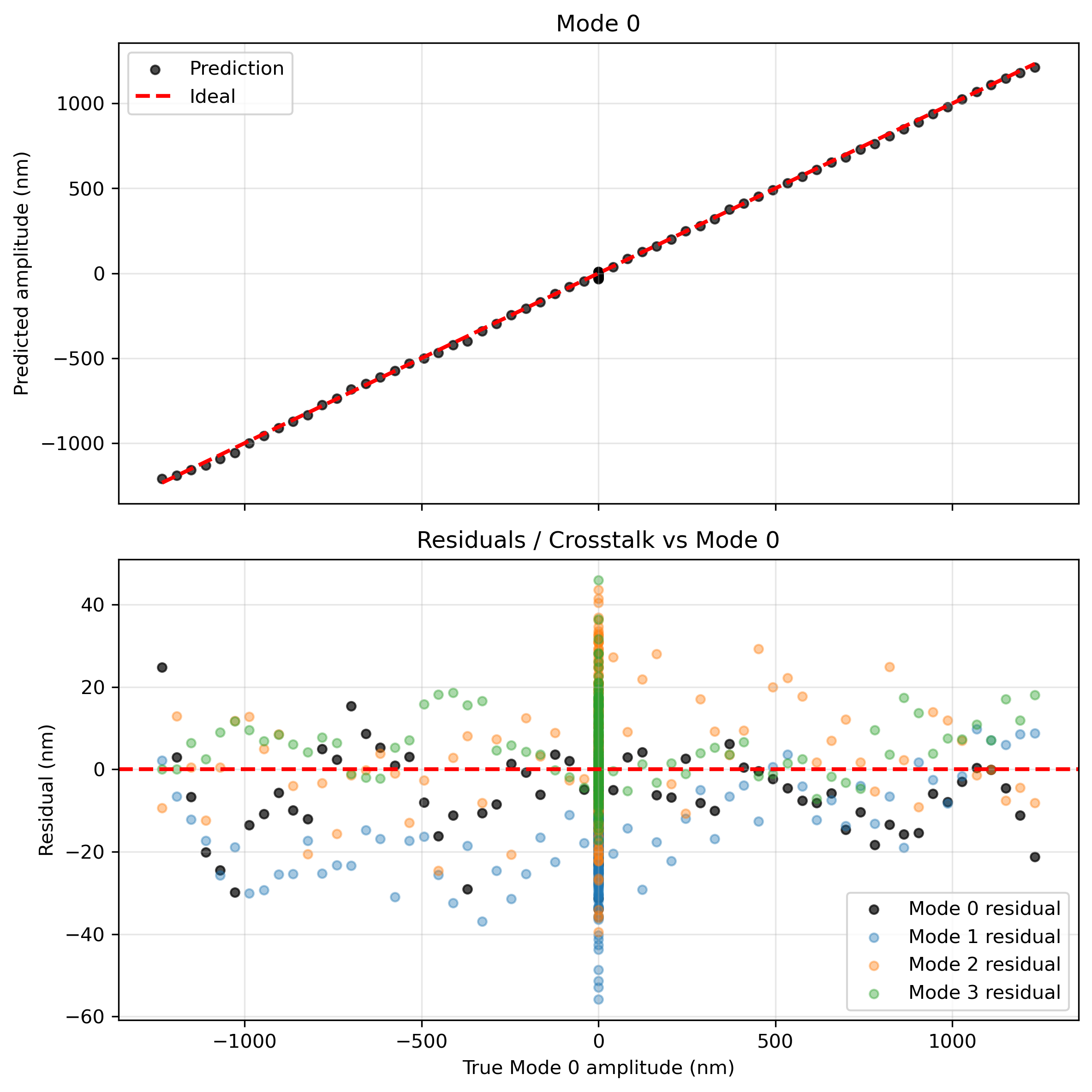}
\end{tabular}
\end{center}
\caption[example] 
{\label{fig:lin} Linearity plot of a piston mode. The x- and the y-axes in the top panel show the amplitude of the piston mode applied to the DM (in nm) and the amplitude predicted by the CNN model (reconstructor), respectively. The bottom plot zooms in on the zero point of the y-axis to show the residual of the mode under study (black) and crosstalk with the three other pupil piston modes (in blue, orange, and green). The non-zero residuals in all the modes when no aberrations were applied could be due to detector noise.}
\end{figure}

For the Linearity test with ML-LLOWFS, known differential piston offsets were sequentially introduced between the four pupil quadrants using the DM. The amplitude of each mode was scanned over a range of $\pm$5~radians, and the corresponding LLOWFS images were recorded. We then used the CNN model computed in Sect.~\ref{sec:cnn-arch} as a reconstructor to predict the amplitude values for each mode applied. Figure~\ref{fig:lin} presents the linearity plot of a piston mode applied in a quadrant of the pupil. The top panel shows the amplitude of the mode applied to the DM (in nm) on the x-axis and the amplitude values predicted (black) by the network on the y-axis. The dashed red line shows the ideal case where y = x. The sensor response appears linear for the entire amplitude range scanned. The bottom panel is a zoom of the top panel around the zero point of the y-axis. This plot shows in black data points the residual in the prediction of the mode under study and cross-talk with the other modes, all within $\pm$40~nm. The non-zero prediction in the reference, when no aberration was applied to the DM, could be due to detector noise and should decrease if the source flux is increased. We will investigate the cause of the non-zero residual at the zero point in future work.    

For the on-sky tests, we observed Altair during the engineering run on 2026-07-03. We built a library with the known pupil-quadrant piston modes and the corresponding LLOWFS images, and computed the CNN model a few hours before the observing run. We then used this CNN model to predict the piston modes for on-sky LLOWFS images. The upstream AO3K and SCExAO loops with their corresponding NIR and visible Pyramid wavefront sensors were closed, respectively. The ML-LLOWFS loop was closed at 1kHz speed with a gain of 0.05 and a leak of 0.89. SCExAO's visible Pyramid wavefront sensor changed its zero point when ML-LLOWFS sent its correction command to SCExAO's DM. 

\begin{figure} [ht]
\begin{center}
\begin{tabular}{c} 
\includegraphics[height=12cm]{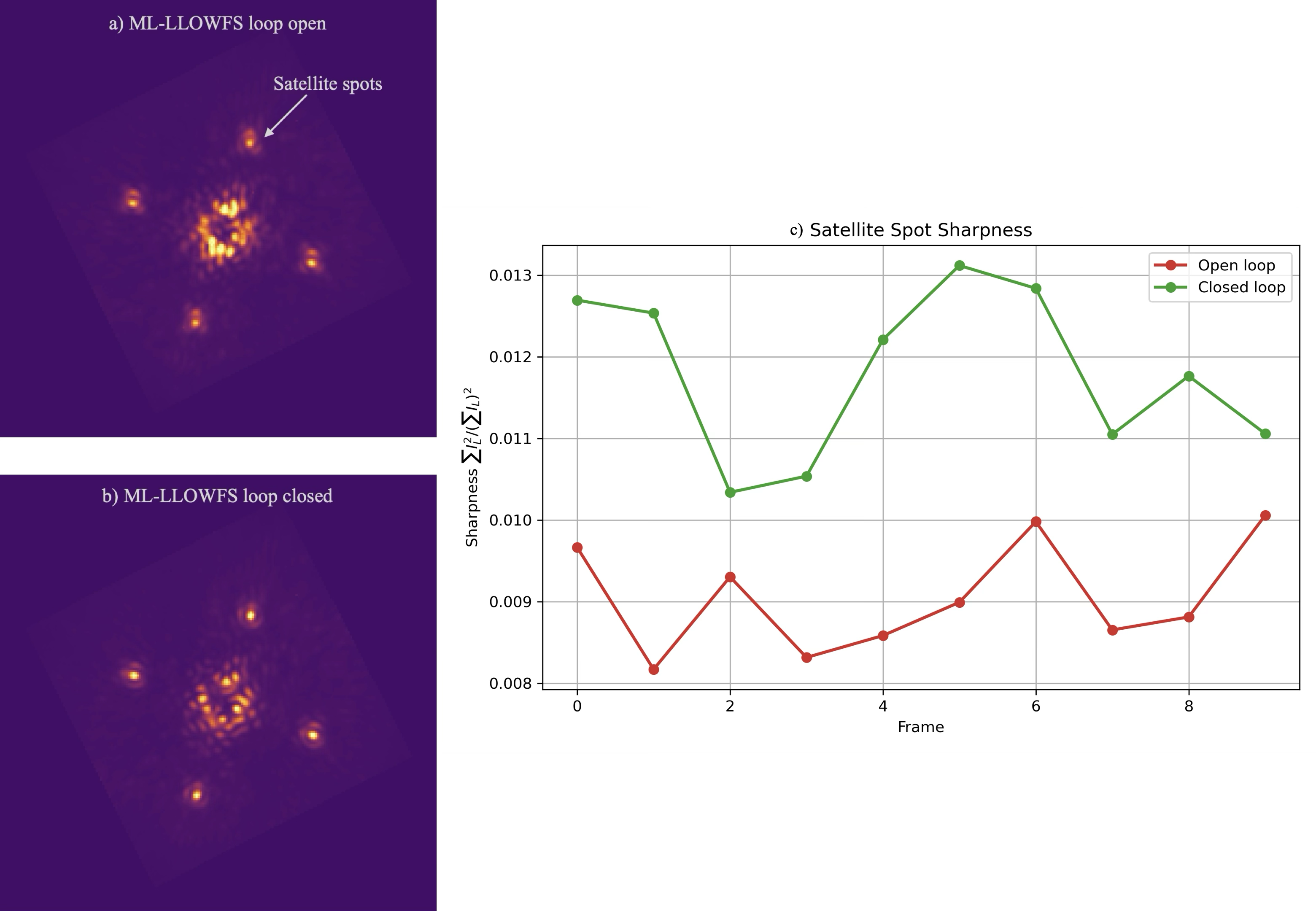}
\end{tabular}
\end{center}
\caption[example] 
{\label{fig:on-sky} The on-sky images on the left show the average of ten 20-second CHARIS frames in the H-band on Altair during the July 2026 engineering run. The top image shows the ML-LLOWFS open-loop with splitting in the satellite spots due to LWE, while the bottom image is with closed-loop showing better PSF morphology of the spots and better temporal stability. The starlight leaks around the FPM due to LWE are more apparent in the open-loop image, with leaked residuals extended diagonally. Both AO3K and SCExAO's NIR and visible Pyramid wavefront sensors were closed, respectively. Both images have the same intensity scale. The plot on the right shows the sharpness of the top-right satellite spot for the open- and closed-loop scenario for ten CHARIS frames. Compared to the open-loop images (red), the spots are compact and well-focused in closed-loop images (green) due to the correction of the LWE.}
\end{figure}

Figure~\ref{fig:on-sky} shows an average of ten 20-second CHARIS frames in the H-band, with the open-loop image displayed on the top-left and the closed-loop image on the bottom-left using the same intensity scale. The sinusoidal phase pattern with an amplitude of 50~nm at $\approx11~\lambda/D$ is applied to the DM to create artificial diffraction spots for accurate astrometry and photometric calibration. These four satellite spots are the best indicator of the LWE impact on the coronagraphic images. The spots in the open-loop image are split due to LWE, and there is more starlight leakage around the FPM, extending diagonally; however, the same spots look like diffraction copies of the stellar point spread function (PSF), with comparatively less starlight leakage around the mask in closed-loop. Note that the ML-LLOWFS loop was closed only on the piston modes, and not on tip-tilt, which will be included in future tests. The first Airy ring of satellite spots in the closed-loop image still shows some splitting, but the core of the spots is mostly intact. There were some stability issues during closed-loop, which we are currently investigating. 

The image on the right in Fig.~\ref{fig:on-sky} shows the sharpness of the satellite spot per frame, computed using the intensity-squared metric, which measures how concentrated the light is within the selected Region of Interest (top right spot) in the averaged open (Fig.~\ref{fig:on-sky}(a)) and closed-loop (Fig.~\ref{fig:on-sky}(b)) images. In the plot, the closed-loop images show higher spot sharpness as the light is concentrated in a small number of pixels, indicating a compact, well-focused spot compared to the open-loop images. These preliminary results are encouraging, and with enhanced loop stability, ML-LLOWFS is envisioned for regular use during SCExAO's science operations to correct the impact of LWE in the coronagraphic mode.

\section{Conclusions}
\label{sec:conclusion}  
Following the on-sky success of Machine Learning with LLOWFS (ML-LLOWFS) on the SPIDERS instrument\cite{fogal-ml-llowfs}, we implemented a similar machine learning approach to tackle low-wind modes on the SCExAO instrument of the Subaru Telescope. The Fast and Furious technique on SCExAO has been used to correct the low-wind modes on-sky in the non-coronagraphic mode (Vievard in prep. \cite {FnF}). However, it has not been successful yet in the coronagraphic mode during SCExAO's on-sky operations. In this article, we presented the preliminary on-sky correction of the differential piston aberrations induced by the LWE using ML-LLOWFS on the post-AO3K and SCExAO wavefront residuals. For our tests, we used a focal-plane mask with an IWA of 113~mas. The loop was closed on the four pupil-quadrant piston modes on Altair during an engineering run in July 2026, demonstrating improved temporal stability. Future work includes diagnosing the loop stability issue, closing the loop on tip-tilt and Zernike modes in addition to the differential piston, closing the ML-LLOWFS loop in conjunction with the Electric Field Conjugation\cite{ao3k-lozi}, adding ML-LLOWFS data in the PSF calibration using wavefront sensor telemetry\cite{psf-telem} on SCExAO, and designing a new focal-plane mask that directs most unused starlight towards the LLOWFS path for better sensitivity of the sensing signal.

\acknowledgments 
 
The development of SCExAO and AO3k is supported by the Japan Society for the Promotion of Science (Grant-in-Aid for Research $\#$23340051, $\#$26220704, $\#$23103002, $\#$19H00703, $\#$19H00695, and $\#$21H04998), Subaru Telescope, the National Astronomical Observatory of Japan, the Astrobiology Center of the National Institutes of Natural Sciences, Japan, the Mt Cuba Foundation, and the Heising-Simons Foundation. The development of the CACAO software is supported by the National Science Foundation under the award $\#$2410616. The authors wish to recognize and acknowledge the very significant cultural role and reverence that the summit of Maunakea has always had within the indigenous Hawaiian community, and are most fortunate to have the opportunity to conduct observations from this mountain. GS gratefully acknowledges the Subaru Telescope Day Crew staff for their daily instrument checks and telescope operators for nighttime support, without which the nighttime observations would not have been possible. GS would also like to thank ChatGPT for advice on building an effective LLOWFS library. 

\bibliography{bib_GS} 
\bibliographystyle{spiebib} 

\end{document}